\documentclass[letterpaper]{article}
\usepackage{arxiv1106} 
\usepackage{times} 
\usepackage{helvet} 
\usepackage{courier} 
\usepackage[hyphens]{url} 
\usepackage{graphicx}
\usepackage{natbib}
\usepackage{caption}
\usepackage{algorithm}
\usepackage{algorithmic}

\usepackage{amsmath}
\usepackage{multirow}
\usepackage{booktabs}
\usepackage{tabularx}
\usepackage{soul}

\nocopyright

\title{Fine-Grained Guidance for Retrievers: Leveraging LLMs' Feedback in Retrieval-Augmented Generation}
\author{
    Yuhang Liu\textsuperscript{\rm 1},
    Xueyu Hu\textsuperscript{\rm 1},
    Shengyu Zhang\textsuperscript{\rm 1,\thanks{Corresponding author: sy\_zhang@zju.edu.cn}},
    Jingyuan Chen\textsuperscript{\rm 1},
    Fan Wu\textsuperscript{\rm 2},
    Fei Wu\textsuperscript{\rm 1}
}
\affiliations{
    \textsuperscript{\rm 1}Zhejiang University, \textsuperscript{\rm 2}Shanghai Jiao Tong University
}

\usepackage{bibentry}

\begin{document}

\maketitle

\begin{abstract}
Retrieval-Augmented Generation (RAG) has proven to be an effective method for mitigating hallucination issues inherent in large language models (LLMs). Previous approaches typically train retrievers based on semantic similarity, lacking optimization for RAG. More recent works have proposed aligning retrievers with the preference signals of LLMs. However, these preference signals are often difficult for dense retrievers, which typically have weaker language capabilities, to understand and learn effectively.
Drawing inspiration from pedagogical theories like Guided Discovery Learning, we propose a novel framework, FiGRet (Fine-grained Guidance for Retrievers), which leverages the language capabilities of LLMs to construct examples from a more granular, information-centric perspective to guide the learning of retrievers.
Specifically, our method utilizes LLMs to construct easy-to-understand examples from samples where the retriever performs poorly, focusing on three learning objectives highly relevant to the RAG scenario: relevance, comprehensiveness, and purity. These examples serve as scaffolding to ultimately align the retriever with the LLM's preferences. Furthermore, we employ a dual curriculum learning strategy and leverage the reciprocal feedback between LLM and retriever to further enhance the performance of the RAG system.
A series of experiments demonstrate that our proposed framework enhances the performance of RAG systems equipped with different retrievers and is applicable to various LLMs.
\end{abstract}

\section{Introduction}
Large language models (LLMs), such as GPT-4 \citep{achiam2023gpt}, have achieved impressive results across a wide range of language tasks \citep{brown2020language, kojima2022large}.
However, despite their rapid recent development, the issue of hallucinations persists \citep{zhang2023siren, ji2023survey}.
Particularly in knowledge-intensive tasks \citep{kandpal2023large}, the generated content sometimes deviates from factual information, resulting in fabricated or inaccurate statements.

Retrieval-Augmented Generation (RAG) is regarded as an effective approach to mitigate the issue of hallucination \citep{lewis2020retrieval, borgeaud2022improving}, by leveraging external corpora to assist LLMs in generating accurate factual information.

In previous approaches, retrievers are typically trained on semantic similarity, which may not align well with LLMs.
Recent efforts \citep{shi2023replug, yu2023augmentation} have explored fine-tuning retrievers with the results generated by LLMs as supervision signals, often involving LLMs annotating preferred documents within the corpus.
However, due to their weaker linguistic abilities, retrievers struggle to grasp the fine-grained preferences of LLMs, posing challenges for effective alignment.
Drawing from educational theories such as discovery learning, we note that training retrievers directly on LLMs' preferences resembles pure discovery learning, which lacks explicit hints about why certain documents benefit LLMs. Literature on this theory \citep{craig1956directed, kittell1957experimental, piaget1970science, mayer2004should} suggests that guided discovery—providing extra guidance and feedback—is often more effective than pure discovery for learning concepts and rules.

Therefore, we propose the FiGRet framework, inspired by educational theories. We view the more linguistically capable LLM as the 'teacher' and the smaller model (retriever) as the 'student'. We provide feedback and guidance on the retriever's alignment through the four steps commonly found in guided discovery learning scenarios, as well as three objectives closely related to RAG performance.

Specifically, the proposed framework follows four steps: establishing learning objectives, constructing guidance, student model training, and assessing performance. 

First, we identify three key, minimally overlapping factors affecting RAG performance as learning objectives: 1) \textbf{Relevance:} Retrieved documents must contain directly relevant information to ensure the LLMs can generate correct results. 2) \textbf{Comprehensiveness:} The completeness of information within retrieved documents influences the comprehensiveness of the LLM's generated content. 3) \textbf{Purity:} The proportion of noise (irrelevant information) within a document impacts generation quality, as excessive noise can mislead the model or obscure relevant information. We refer to the proportion of non-noisy information within a document as its purity. These objectives guide the retriever towards our ultimate goal of aligning with LLMs preferences.

Second, we guide the retriever to learn these objectives through a more granular perspective. While previous work focused on training at the document-level, the retriever already possesses substantial document-level understanding. To further enhance its capabilities, we shift from a document-centric to an information-centric perspective, viewing each document as a collection of information units. Our learning objectives can then be approximated as the accuracy, recall, and precision of all retrieved information units. Following common practices in guided discovery learning, we construct easy-to-understand examples with hints across these three objectives. This construction facilitates the less linguistically capable retriever in capturing and aligning with the LLMs' complex preferences without relying on implicit learning from massive document pairs.

Third, we adopt a dual curriculum learning approach for student models training, gradually increasing the difficulty of learning tasks as suggested in guided discovery practices \citep{shulman1966learning, mayer2004should}.

Finally, we conduct performance assessment, and leverage well-learned and poorly-learned instances to further enhance the framework. Well-learned instances are used to optimize the teacher model's guidance construction, while poorly-learned instances undergo additional learning.

Our experiments show that various retrievers achieve performance improvements within our framework. Across tasks such as MMLU and open-domain QA, performance improvements are observed across different LLMs. Furthermore, we validate the learning effectiveness of the retrievers in the three objectives. We summarize our contributions as follows:

\begin{itemize}
    \item We propose a framework, FiGRet, inspired by educational theories, in which LLMs assist smaller models (retrievers) in learning by providing high-quality guidance, enabling the smaller models to more efficiently learn complex knowledge, such as LLMs' preferences.
    \item We construct guidance examples based on three key factors affecting RAG performance, adopting a fine-grained perspective to help retrievers align with LLMs' preferences.
    \item Our framework allows for feedback and guidance from black-box LLMs without needing access to their inference processes, simplifying deployment.
\end{itemize}

\section{Related Work}

\paragraph{Retrieval-Augmented Generation}
RAG aims to enhance language models by retrieving and integrating relevant information from external knowledge sources, demonstrating significant improvements in handling knowledge-intensive tasks \citep{lewis2020retrieval, pmlr-v119-guu20a}. RAG models excel in various NLP applications, like question answering \citep{izacard2021leveraging, 10.1162/tacl_a_00605} and summarization \citep{lin2023ra}.

Several retrieval models have been developed to support RAG frameworks. BM25 \citep{robertson2009probabilistic}, a sparse retrieval model, has been a foundational method for text retrieval tasks. DPR \citep{karpukhin2020dense} indexes passages into a dense vector space for efficient retrieval and has been widely adopted in subsequent RAG models. Other notable dense retrieval models include Contriever \citep{izacard2021unsupervised}, SBERT \citep{reimers2019sentence}, and BGE \citep{xiao2023bge}, each contributing to the robustness and effectiveness of retrieval-augmented systems.

Recent advancements in retrieval strategies have optimized the interaction between retrievers and language models. Models like Atlas \citep{izacard2023atlas} and RETRO \citep{borgeaud2022improving} employ joint training and architectural modifications to better integrate retrieved information, though these methods are resource-intensive.

\paragraph{LLM-Supervised Learning}
LLMs have demonstrated remarkable capabilities in natural language understanding and generation \citep{brown2020language, chowdhery2022palm}.
Leveraging the knowledge and capabilities of LLMs to guide the training of other models or themselves has recently emerged as a promising direction.
Reinforcement Learning from AI Feedback (RLAIF) \citep{lee2023rlaif} proposes using an LLM to generate preference labels to train a reward model, which then guides the reinforcement learning process, achieving performance comparable to traditional human feedback-based approaches.

In addition, \citet{wang2022self} propose an LLM-bootstrapping approach called self-instruct, where an LLM iteratively generates additional training data for itself. \citet{wang2023learning} introduce the LLM-R framework which trains dense retrievers to identify high-quality in-context examples using feedback from LLMs.

Several recent works \citep{shi2023replug, yu2023augmentation, zhang2023retrieve} leverage the preferences of large models as supervision to train retrievers. For example, the Augmentation-Adapted Retriever (AAR, \citet{yu2023augmentation}) trains a retriever using preferences obtained from a small source LM, which can then enhance the zero-shot generalization of larger target LMs by retrieving relevant documents without requiring fine-tuning of the LMs.
Our method takes a different approach by not directly training the retrieval model based on LM preferences. Instead, we leverage the language capabilities of LLMs to construct high-quality guidance that enables the retriever to learn more effectively.

\section{Preliminaries}

The RAG process typically involves augmenting the input query $x$ with relevant documents $\mathcal{D}$ from a predefined corpus $\mathcal{C}$. The output $y$ can then be defined as:
\begin{equation}
y = \operatorname{LLM_{\theta}}(x \oplus \mathcal{D}), \quad \mathcal{D} \subseteq \mathcal{C}
\end{equation}
The retriever functions by taking an input query and searching the corpus to locate relevant documents. Recent advancements in RAG have focused primarily on the dense retriever, which utilize a pre-trained encoder $\text{Enc}_{\phi}$, commonly a transformer-based model with parameters $\phi$, to embed both the documents in the corpus and the incoming queries into a shared vector space:
\begin{equation}
\mathbf{x} = \operatorname{Enc_{\phi}}(x)
\end{equation}
\begin{equation}
\mathbf{d} = \operatorname{Enc_{\phi}}(d), \quad d \in \mathcal{C}
\end{equation}
where $\mathbf{x}$ and $\mathbf{d}$ represent the query and document embeddings, respectively.

Document retrieval is performed by identifying the document embeddings that are nearest to the query embedding within the vector space. This is usually achieved by computing the dot product between the query embedding and each document embedding:
\begin{equation}
\operatorname{sim}(x, d_i) = \mathbf{x} \cdot \mathbf{d}_i, \quad \forall i \in \{1, 2, \ldots, n\}
\end{equation}

Efficient retrieval is often facilitated by utilizing methods such as approximate nearest neighbor (ANN) search \citep{10.14778/2856318.2856324, johnson2019billion}. The top-$k$ documents with the highest relevance scores are then retrieved:
\begin{equation}
\begin{split}
\mathcal{D}_{\text{ret}} = \operatorname{top}_k(&\operatorname{sim}(x, d_1), \\
&\operatorname{sim}(x, d_2), \ldots, \operatorname{sim}(x, d_n))
\end{split}
\label{eq:retrieve}
\end{equation}

The training of the encoder is typically conducted using a triplet-based approach. This involves training on triplet data consisting of an input query $x$, positive documents $\mathcal{D}^+$, and negative documents $\mathcal{D}^-$:
\begin{equation}
(x, \mathcal{D}^+, \mathcal{D}^-)
\end{equation}

\begin{figure*}[ht]
\centering
\includegraphics[width=\textwidth]{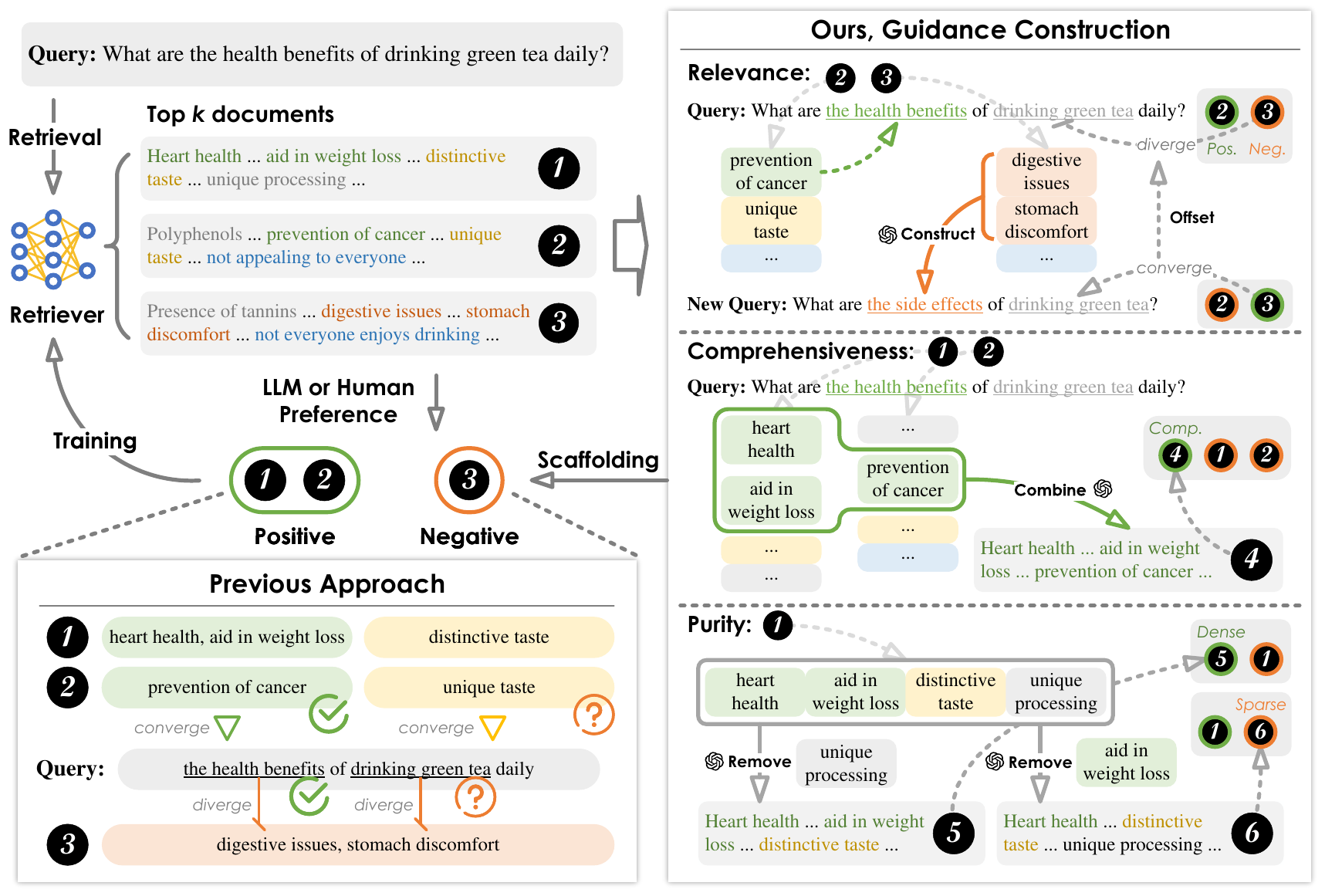}
\caption{
An illustration of the guided construction process within the FiGRet framework. The colored blocks represent the perspective of treating documents as collections of information units from an information-centric perspective. The top-left corner depicts the retrieval and training process of a typical RAG system. The bottom-left corner illustrates the previous approach, which considers each document as a monolithic unit, leading to bias due to the erroneous convergence or divergence of certain information units mistakenly deemed relevant or irrelevant (indicated by the red question marks). Our approach, depicted on the right, mitigates this bias by constructing guided examples through an information-centric perspective, facilitating targeted learning towards specific objectives. In this depiction, "Pos." represents Positive, "Neg." represents Negative, and "Comp." represents Comprehensive.
}
\label{fig:method}
\end{figure*}

\section{Fine-Grained Guidance for Retrievers}\label{subsec:framework}
In this section, we will introduce the specific details of the FiGRet framework.

\subsection{Establishing Learning Objectives}
Our ultimate goal is to align the retriever with the preferences of LLMs, enabling it to identify documents most likely to facilitate optimal LLMs generation given a user query.
Therefore, our framework first establishes explicit and relatively easy-to-learn objectives that serve as scaffolding towards the ultimate goal. 

These objectives are carefully selected based on three key criteria: 1) clarity and interpretability by the teacher model for effective guidance construction, 2) strong positive correlation with the alignment goal, and 3) minimal overlap to avoid redundant guidance. Based on these criteria, we focus on relevance, comprehensiveness, and purity, each addressing a distinct aspect of document quality:

\begin{itemize}
\item \textbf{Relevance}: This objective emphasizes the fundamental importance of content relevance between the document and the query. 
Given that retrievers typically possess a strong foundational ability in document-level relevance, 
our guidance aims to refine and enhance this existing capability.
\item \textbf{Comprehensiveness}: This objective emphasizes the importance of retrieving documents containing rich information related to the query. Our framework guides retrievers to prioritize documents exhibiting higher levels of comprehensiveness.
\item \textbf{Purity}: This objective emphasizes the detrimental impact of noisy information within documents on LLMs generation. Our guidance aims to steer retrievers towards documents with minimized noise.
\end{itemize}

We then leverage the teacher LLM, denoted as $\text{LLM}_T$, to assess the retriever's mastery of each learning objective.
Recognizing the foundational role of relevance in RAG, the teacher primarily focuses on query-document relevance, creating three broad scoring categories. Within each category, distinctions are made based on comprehensiveness and purity. To ensure consistent and accurate evaluation, the top-$k$ retrieved documents for a given query are scored collectively using chain-of-thought \citep{wei2022chain} prompting. Finally, to mitigate potential inconsistencies in scoring across different queries, we convert scores to rankings and utilize normalized discounted cumulative gain (NDCG, \citet{jarvelin2002cumulated}) to measure the discrepancy between the retriever's ranking and the ideal ranking. 
This ranking-based NDCG score serves as a proxy for the retriever's mastery of the learning objectives.

\subsection{Constructing Guidance}
Guidance construction is tailored to the retriever's mastery of the learning objective. During online inference, we collect input queries and the retriever's top-$k$ retrieved documents. When the system is idle, we utilize the NDCG metric to evaluate these samples and populate a sample pool. Once the pool reaches a predefined size, we empirically determine an NDCG threshold for guidance selection. This threshold is set as the minimum NDCG score among samples where the top-1 ranking given by the LLM matches the retriever's ranking. Samples falling below this threshold are selected for guidance.

The construction process of the guidance examples is as follows, and is illustrated in Figure~\ref{fig:method}.

\paragraph{Relevance.}
In this objective, we construct guidance examples by reversing the positive and negative documents for the input query $x$. By doing so, we enable the retriever to focus on more granular levels of relevance. Specifically, in the previous document-centric approach, during training, the information units within the negative documents $\mathcal{D}^-$ are made to diverge from all units in query $x$, including those that are inherently relevant. This introduces a bias and can be viewed as a reduction in the accuracy of information units, as they are incorrectly deemed irrelevant to the query (illustrated in the bottom-left corner of Figure~\ref{fig:method}). To mitigate this, for the information units in $\mathcal{D}^-$ that are irrelevant to $x$, we construct a new query $x'$, which only changes the relevance from irrelevant to relevant with respect to the units in $\mathcal{D}^-$.

The specific construction approach is as follows: We first extract the information differences between $ \mathcal{D}^+ $ and $ \mathcal{D}^- $.
\begin{equation}
\begin{aligned}
\Delta_i &= \operatorname{Info}(\mathcal{D}^-_i) - (\operatorname{Info}(\mathcal{D}^{+}) \cap \operatorname{Info}(\mathcal{D}^-_i)) \\
         &= \operatorname{LLM}_T(\mathcal{D}^+, \mathcal{D}^-)
\end{aligned}
\end{equation}
where $ \mathcal{D}^-_i \subseteq \mathcal{D}^-$ is the subset containing content-similar $d^-$. Note that $\operatorname{Info}()$ represents viewing documents as sets of information units, so $(\operatorname{Info}(\mathcal{D}^{+}) \cap \operatorname{Info}(\mathcal{D}^-_i))$ represents the shared information between $\mathcal{D}^+$ and $\mathcal{D}^-_i$. Thus, $\Delta_i$ represents the information present in $ \mathcal{D}^-_i$ but absent in $\mathcal{D}^+$.

Next, we construct a new input $x'_i$ corresponding to $\mathcal{D}^-_i$ by applying $\operatorname{LLM}_T$ to $\Delta_i$.

For $x'_i$, $\mathcal{D}^-_i$ is the relevant document set, while $\mathcal{D}^+$ and $\mathcal{D}^- - \mathcal{D}^-_i$ (i.e., $\mathcal{D}^N_{\text{ret}} - \mathcal{D}^-_i$) are not relevant. Therefore, we obtain the following examples:
\begin{equation}
\left\{
\begin{aligned}
&\left( x, \mathcal{D}^+, \mathcal{D}^- \right); \\
&\left( x'_i, \mathcal{D}^-_i, \mathcal{D}^k_{\text{ret}} - \mathcal{D}^-_i \right), \forall i
\end{aligned}
\right\}
\end{equation}

\paragraph{Comprehensiveness.} 
In this objective, to construct guiding examples that facilitate the retrieval of more comprehensive documents, the most intuitive approach is to either enhance or reduce the comprehensiveness of a document to create a contrast. In this case, we have developed a new comprehensive document containing more information. 

Specifically, we construct a new, more comprehensive document, denoted as $d_{\text{comp}}$, by leveraging all relevant information from $\mathcal{D}^+$ and the parameter knowledge of $\operatorname{LLM}_T$. This document is generated by applying $\operatorname{LLM}_T$ to the combined information:

\begin{equation}
d_{\text{comp}} = \operatorname{LLM}_T\left( \operatorname{Info^+}(\mathcal{D}^+), \theta \right)
\end{equation}

where $\operatorname{Info^+}(\mathcal{D}^+)$ represents the relevant information contained in $\mathcal{D}^+$, and $\theta$ represents the the parameter knowledge of $\operatorname{LLM}_T$. Then, we can derive the following example:
\begin{equation}
\left\{
(x, d_{\text{comp}}, \mathcal{D}^+)
\right\}
\end{equation}

This setup allows the retriever to recognize documents that are more comprehensive.

\paragraph{Purity} 
In this objective, we guide the retriever to identify documents with a lower proportion of noisy information by modifying the ratio of noisy information units within the documents.

Specifically, we first use the $\operatorname{LLM}_T$ to remove noisy and non-noisy information units from $d^+$, respectively, thereby altering the document's purity:
\begin{equation}
\begin{aligned}
d_{\text{dense}} &= \operatorname{LLM}(\operatorname{Info}(d^+) - \{u^- | u^- \in \operatorname{Info}(d^+)), \\
d_{\text{sparse}} &=\operatorname{LLM}(\operatorname{Info}(d^+) - \{u^+ | u^+ \in \operatorname{Info}(d^+))
\end{aligned}
\end{equation}
where $u^+$ and $u^-$ denote non-noisy and noisy information units, respectively. This process leads to the construction of two types of examples for each $d^+$ in $\mathcal{D}^+$:
\begin{equation}
\{(x, d_{\text{dense}}, d^+), \quad (x, d^+, d_{\text{sparse}})\}
\end{equation}

\subsection{Student Model Learning}\label{subsec:training}

We employ a dual curriculum learning approach to train the student model. Through the preceding steps, we obtain two sets of data: guidance examples constructed by the teacher model and preference data formed by the teacher model's scoring. We design the training process such that the guidance examples, serving as scaffolding, have a higher distribution in the early stages of training, while the proportion of preference data gradually increases. The probability distribution satisfies the following equation:

\begin{equation}
P_1(i) = \frac{e^{(2y_i-1)/T_1}}{\sum_{j=1}^{N} e^{(2y_j-1)/T_1}}
\end{equation}
where $y_i$ is the binary label of the $i$-th sample (1 for guidance examples and 0 for preference data), and $T_1$ is the temperature parameter.

Simultaneously, we control the difficulty of the training samples to gradually increase based on their NDCG scores, following the distribution:

\begin{equation}
P_2(i) = \frac{e^{s_i/T_2}}{\sum_{j=1}^{N} e^{s_j/T_2}}
\end{equation}
where $s_i$ represents the NDCG score of the $i$-th sample, and $T_2$ is another temperature parameter.

Through this dual curriculum learning approach, we enable the training process to transition from learning the scaffolding of the objectives to the final goal of aligning with the LLMs' preferences while ensuring a gradual increase in learning difficulty.

Then, following previous work \citep{xiong2020approximate}, the training loss $ L $ is compute as follows:
\begin{multline}
L = \sum^{x} \sum_{d^+ \in D^+} \sum_{d^- \in D^-} \\
  l(\operatorname{sim}(x, d^+), \operatorname{sim}(x, d^-))
\end{multline}
where $ l $ denotes the standard cross-entropy loss.

\subsection{Assessing Performance}
After the learning process, we re-evaluate the difficult samples. If a sample's NDCG score surpasses a predefined threshold, we consider it a well-learned sample and use it as a teaching case for the teacher model. During the construction of similar samples, these well-learned samples are provided to the teacher model through retrieval as exemplars for in-context learning (ICL). Conversely, samples whose scores have decreased will undergo additional guidance.

\section{Experiments}

\begin{table*}[ht]
\centering
\small
\begin{tabular}{lcccccccccc}
\toprule
\midrule
\multirow{2}{*}{\textbf{Method}} & \multicolumn{5}{c}{\textbf{MMLU}} & \multirow{2}{*}{\textbf{NQ}} & \multirow{2}{*}{\textbf{PQA}} & \multirow{2}{*}{\textbf{HoPo}} & \multirow{2}{*}{\textbf{FEV.}} & \multirow{2}{*}{\textbf{All Avg.}} \\ \cline{2-6}
 & Hum. & Soc. & STEM & Other & All &  &  &  &  &  \\ 
\midrule
\multicolumn{11}{l}{\textit{GPT-3.5-Turbo}} \\ 
\midrule
No retrieval & 52.9 & 76.6 & 53.1 & 75.7 & 63.4 & 48.1 & 44.3 & 33.6 & 82.1 & 54.3 \\
Contriever & 55.1 & 76.3 & \ul{54.5} & 74.5 & 64.2 & 48.8 & 45.6 & 39.0 & 89.4 & 57.4 \\
AAR$_{\text{Contriever}}$ & 54.3 & \ul{78.5} & 52.5 & \ul{77.1} & 64.4 & 49.0 & 46.3 & 36.9 & 89.6 & 57.2 \\
BGE & 52.9 & 78.2 & 54.0 & 76.5 & 64.1 & \ul{50.3} & 43.9 & 39.5 & 89.3 & 57.4 \\
SBERT & 54.1 & 77.9 & 52.8 & \textbf{77.4} & 64.5 & 49.4 & \textbf{50.1} & 38.7 & 88.5 & 58.2 \\ 
\midrule
FiGRet$_{\text{Contriever}}$ (Ours) & \ul{55.4} & 76.9 & \ul{54.5} & \ul{77.1} & 65.0 & 49.6 & \ul{48.0} & \ul{39.9} & \textbf{90.6} & \textbf{58.6} \\
FiGRet$_{\text{BGE}}$ (Ours) & \textbf{55.8} & \textbf{79.8} & 54.3 & 76.5 & \textbf{65.5} & \textbf{50.4} & 45.7 & \textbf{40.0} & \ul{90.3} & \ul{58.4} \\
FiGRet$_{\text{SBERT}}$ (Ours) & \textbf{55.8} & 77.2 & \textbf{55.2} & 76.8 & \ul{65.4} & 49.9 & \textbf{50.1} & 39.1 & 88.7 & \textbf{58.6} \\ 
\midrule
\multicolumn{11}{l}{\textit{Llama-3-8B-Instruct}} \\
\midrule
No retrieval & 52.9 & 74.4 & 51.6 & 73.3 & 62.0 & 33.1 & 26.1 & 25.9 & 79.1 & 45.2 \\
Contriever & 52.9 & 76.3 & 52.5 & 73.3 & 62.5 & 41.3 & 41.7 & 36.0 & 84.5 & 53.2 \\
AAR$_{\text{Contriever}}$ & 52.9 & \ul{77.2} & \ul{54.0} & 73.9 & 63.2 & 42.1 & 42.3 & 35.3 & 85.2 & 53.6 \\
BGE & \ul{54.4} & 76.9 & 52.8 & 73.6 & 63.3 & \ul{44.1} & 36.1 & 35.9 & 86.1 & 53.1 \\
SBERT & 53.9 & 76.6 & \textbf{54.6} & 73.3 & 63.4 & 41.7 & \ul{46.0} & 35.7 & 86.2 & 54.6 \\ 
\midrule
FiGRet$_{\text{Contriever}}$ (Ours) & 53.5 & \textbf{77.5} & 53.1 & \ul{74.8} & 63.4 & 43.0 & 44.3 & \ul{36.9} & \ul{86.5} & 54.8 \\
FiGRet$_{\text{BGE}}$ (Ours) & 53.9 & 76.3 & 53.4 & \textbf{75.1} & \ul{63.6} & \textbf{45.3} & 41.4 & \textbf{37.5} & \textbf{87.8} & \textbf{55.1} \\
FiGRet$_{\text{SBERT}}$ (Ours) & \textbf{54.6} & 76.3 & \ul{54.0} & 74.2 & \textbf{63.7} & 42.8 & \textbf{46.3} & 36.1 & 86.2 & \ul{55.0} \\ 
\midrule
\multicolumn{11}{l}{\textit{Claude-3-Haiku}} \\ 
\midrule
No retrieval & 59.5 & 82.6 & 59.4 & 78.3 & 68.8 & 27.6 & 31.7 & 26.9 & 70.4 & 45.1 \\
Contriever & 61.6 & 82.0 & \ul{60.0} & \ul{79.8} & 70.0 & 35.7 & 41.3 & 33.0 & 90.0 & 54.0 \\
AAR$_{\text{Contriever}}$ & 62.6 & \ul{83.5} & 59.1 & 79.5 & \ul{70.2} & 36.1 & 42.1 & 32.7 & \ul{90.2} & 54.3 \\
BGE & 61.4 & 82.0 & 58.5 & 77.4 & 69.0 & \ul{38.1} & 37.5 & 33.0 & 89.6 & 53.4 \\
SBERT & 62.0 & 81.0 & 58.8 & 79.2 & 69.4 & 35.9 & \ul{46.2} & 32.7 & 89.5 & 54.7 \\ 
\midrule
FiGRet$_{\text{Contriever}}$ (Ours) & 62.9 & 82.0 & \textbf{60.3} & \textbf{80.1} & \textbf{70.5} & 36.5 & 44.2 & \ul{33.8} & \textbf{90.4} & 55.1 \\
FiGRet$_{\text{BGE}}$ (Ours) & \ul{63.3} & 82.9 & 57.9 & 78.9 & 69.9 & \textbf{40.0} & 42.2 & \textbf{35.7} & 90.0 & \textbf{55.6} \\
FiGRet$_{\text{SBERT}}$ (Ours) & \textbf{63.7} & \textbf{84.2} & 59.4 & 77.7 & \textbf{70.5} & 37.1 & \textbf{46.6} & 33.0 & 90.1 & \ul{55.5} \\ 
\midrule
\bottomrule
\end{tabular}
\caption{
Our main results on different tasks. We conducted experiments on three different LLMs. Results are reported for our framework initialized with three different base retrievers. For datasets other than MMLU, we tested using 1,000 samples and employed stratified equidistant sampling to ensure fairness. \textbf{Bold} scores represent the best performance within the same LLM, while \ul{underlined} scores represent the second best.
}
\label{tab: main}
\end{table*}

In this section, we describe the experimental settings used to evaluate our proposed framework.

\subsection{Tasks and Datasets}

We conduct evaluations across the following tasks: Language Understanding, Open Domain Question Answering (Open Domain QA), and Fact Checking. For each task, we introduce the corresponding evaluation datasets:

\paragraph{Language Understanding.} We adopt \textbf{MMLU} \citep{hendrycks2020measuring}, a multiple-choice QA dataset that covers 57 subjects including STEM, humanities, social sciences and others. We report the accuracy on the development split as the metric, following prior work \citep{yu2023augmentation}.

\paragraph{Open Domain QA.} We evaluate on three open domain QA datasets: 1) \textbf{Natural Questions} (NQ; \citet{10.1162/tacl_a_00276}), which contains questions from Google search queries paired with Wikipedia answer passages. We use the open variant from \citet{lee2019latent}; 2) \textbf{HotpotQA} \citep{yang2018hotpotqa}, which features compositional questions requiring reasoning over multiple Wikipedia paragraphs, and we use its fullwiki setting; 3) \textbf{PopQA} \citep{mallen2022not}, which consists of questions about long-tail Wikidata entities with answers extracted from Wikipedia. For this task, we consider a generated answer correct if it contains the gold answer. We report the accuracy on the development split, except for PopQA where the test set was used.

\paragraph{Fact Checking.} We evaluate the model's fact-checking capabilities using the \textbf{FEVER} (Fact Extraction and VERification, \citet{Thorne18Fever}) dataset, which is a large-scale dataset containing claims that are verified against textual evidence from Wikipedia.

We use the KILT version of the dataset from \citet{petroni-etal-2021-kilt}, and report the accuracy on the development split.

For the MMLU tasks, the MSMARCO corpus is used, while Wikipedia\footnote{Wikipedia dump, December 20, 2018 \citep{karpukhin-etal-2020-dense}.} is the corpus for other tasks.

\subsection{Implementation Details}

\paragraph{Teacher LLM.} We utilize GPT-3.5 from OpenAI's GPT family \citep{brown2020language,ouyang2022training} as the teacher LLM due to its strong language abilities and cost-effectiveness. Specifically, we employ \textit{GPT-3.5-Turbo-0125} version.

\paragraph{Runtime Task.} We used \textbf{MSMARCO} \citep{bajaj2016ms} as our runtime task, which contains questions requiring factual knowledge to answer. We randomly shuffled the training set and set the sample pool size to 10,000 samples for each learning iteration. In our experiments, we conducted two iterations. For each sample, we set the top-$k$ document retrieval to 8 for guidance construction.

\paragraph{Training Details.} During training, we used all the constructed guidance examples. For each input $x$, a positive document $d^+$ and a negative document $d^-$ are randomly sampled from $\mathcal{D}^+$ and $\mathcal{D}^-$, respectively, which are used for training. The batch size during training is set to $128$, and the learning rate is set to $5 \times 10^{-6}$. All of our experiments were conducted on 4$\times$A100 (40G) GPUs.

\paragraph{Inference LLMs.} For the inference LLMs, we utilize GPT-3.5, the same as in the guiding stage. Furthermore, to validate the effectiveness of our framework on different LLMs, we test two other widely-used LLMs: Llama-3 \cite{llama3modelcard}, a high-performance open-source language model developed by Meta. We use the \textit{Llama3-8B-Instruct} version, and Claude-3 \cite{anthropic2024claude} from Anthropic, and we employ the \textit{Claude-3-Haiku} version in our experiments. During model generation, we set the temperature to 0 or used greedy search decoding to minimize randomness, and the results were obtained from a single run.

\subsection{Baselines} 

\paragraph{No Retrieval.}
This baseline involves directly generating responses using inference LLMs without the aid of retrieval. In this setting, the LLMs rely solely on their internal parametric knowledge to generate answers.

\paragraph{Retrieval Augmented.}
For the retriever base models, we used Contriever\footnote{\url{https://huggingface.co/facebook/contriever-msmarco}} \citep{izacard2021unsupervised}, BGE\footnote{\url{https://huggingface.co/BAAI/bge-base-en-v1.5}} \citep{xiao2023bge}, and Sentence-BERT\footnote{\url{https://huggingface.co/sentence-transformers/msmarco-bert-base-dot-v5}} \cite{reimers-2019-sentence-bert}, which are competitive and have achieved excellent results in relevant tasks. We also validated the method of aligning retrievers through LLMs preference: AAR, which we adopted the checkpoint\footnote{\url{https://huggingface.co/OpenMatch/AAR-Contriever}} using Contriever as the base model.

We employed a few-shot approach during inference.

\section{Analysis}

In this section, we discuss the main experimental result and present ablation studies.

\subsection{Main Results}\label{sec:main_results}

Our framework is designed to improve the performance of retrievers in RAG systems without requiring external supervision or labeled data. Table~\ref{tab: main} demonstrates that our approach leads to improved LLM generation performance across various LLM and retriever combinations. To demonstrate the performance differences, only documents with the highest similarity scores are utilized. Notably, the performance gains achieved by Contriever within our framework surpass those of AAR, which is trained to align with LLMs' preferences using signals from a smaller source LM.

To validate the generalizability of our approach, we conducted experiments on diverse LLMs. The results reveal consistent performance improvements across these LLMs, highlighting the broad applicability of our method. It is worth noting that we only conducted example construction on 20,000 samples (approximately 1/40 of the original training set) during the training phase, demonstrating the sample efficiency of our framework.

\subsection{Ablation studies}\label{sec:ablation}
\begin{table}[t]
\centering
\begin{tabular}{lcccc}
\toprule
\textbf{Method} & \multicolumn{1}{l}{\textbf{MMLU}} & \multicolumn{1}{l}{\textbf{PQA}} & \multicolumn{1}{l}{\textbf{FEV.}} & \multicolumn{1}{l}{\textbf{Avg.}} \\ 
\midrule
Contriever & 62.5 & 41.7 & 84.5 & 62.9 \\ 
\midrule
w.o. Relevance & \textbf{63.5} & 43.8 & 85.7 & 64.3 \\
w.o. Comp. & 63.2 & 43.0 & 85.7 & 64.0 \\
w.o. Purity & 63.1 & 44.0 & 85.2 & 64.1 \\ 
w.o. Scaffolding & 63.0 & 44.1 & 85.0 & 64.0 \\
w.o. Preference & 63.0 & 43.4 & 85.3 & 63.9 \\
\midrule
Ours & 63.4 & \textbf{44.3} & \textbf{86.5} & \textbf{64.7} \\ 
\bottomrule
\end{tabular}
\caption{
Performance of FiGRet when removing each of the three objectives, respectively. "Comp.” represents comprehensiveness.
}
\label{tab: ablation}
\end{table}

To investigate the impact of each objective on the performance of our framework, we conducted ablation experiments on \textit{Llama-3-8B-Instruct}. Specifically, we removed each of the three objectives individually and evaluated their performance on three diverse tasks: MMLU, PopQA, and FEVER. The results of these experiments are presented in Table~\ref{tab: ablation}.

The ablation study reveals that removing any of the three objectives leads to a certain degree of performance degradation compared to the complete framework. However, it is noteworthy that even with the ablated components, the framework still outperforms the initial retriever.

We also verified the results after removing the guidance examples and preference data. The results show that removing either of these would affect the final performance, demonstrating the connection between learning objectives and the ultimate goal.

Notably, we observed that removing the first objective (i.e. relevance), had the least impact on the framework's performance. We hypothesize that this may be attributed to the model's initially strong understanding of relevance, resulting in lower gains from this objective compared to the others.

\subsection{Evaluating Improvement on Individual Objectives}

\begin{figure}[t]
\centering
\includegraphics[width=0.45\textwidth]{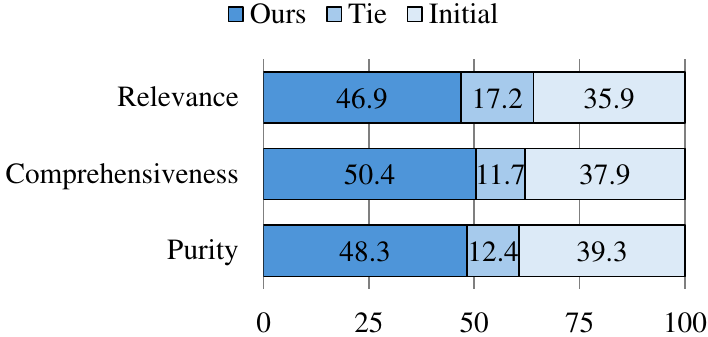}
\caption{
Objective-wise Retrieval Improvements After Training. The retriever demonstrates improved performance across all three objectives
}
\label{fig: 3_aspects}
\end{figure}

To verify whether the retriever has improved on all three objectives after training, i.e., whether these three objectives can be well learned by the model, we compared the retrieved documents by the retriever before and after training on the three objectives. We chose BGE because it achieved the largest gains with our method. We tested 1k samples from the MSMARCO dataset that do not overlap with the training data, and obtained the top 1 documents retrieved by the retriever for the same input before and after training. We removed samples where the top 1 documents were the same, and compared the remaining samples on the three objectives. Specifically, we used GPT-4o to score the two sets of documents on each of the three objectives. To mitigate positional bias, we tested each sample twice with the order of the two documents swapped and compared their average values, considering it a tie when the average values were the same. The results are shown in Figure \ref{fig: 3_aspects}. After applying our framework, the retriever achieved higher win rates on all three objectives than before, demonstrating that the retriever can effectively learn these three objectives.

\section{Conclusion}
We introduce the FiGRet framework, a novel approach for enhancing the alignment between retrievers and LLMs in RAG systems by providing fine-grained feedback and guidance focused on relevance, comprehensiveness, and purity. Our experiments demonstrate performance improvements across various tasks and retrievers when coupled with different LLMs. Notably, our framework obtains feedback directly from black-box LLMs, reducing deployment difficulty, and achieves considerable performance gains with only 20,000 upstream samples.

The FiGRet framework offers a new perspective on enhancing retriever-LLM alignment in RAG systems. Future research could explore integrating the framework with other techniques to further advance RAG systems.

\bibliography{arxiv1106}

\begin{thebibliography}{49}
\providecommand{\natexlab}[1]{#1}

\bibitem[{Achiam et~al.(2023)Achiam, Adler, Agarwal, Ahmad, Akkaya, Aleman, Almeida, Altenschmidt, Altman, Anadkat et~al.}]{achiam2023gpt}
Achiam, J.; Adler, S.; Agarwal, S.; Ahmad, L.; Akkaya, I.; Aleman, F.~L.; Almeida, D.; Altenschmidt, J.; Altman, S.; Anadkat, S.; et~al. 2023.
\newblock Gpt-4 technical report.
\newblock \emph{arXiv preprint arXiv:2303.08774}.

\bibitem[{AI@Meta(2024)}]{llama3modelcard}
AI@Meta. 2024.
\newblock Llama 3 Model Card.

\bibitem[{Andr\'{e}, Kermarrec, and Le~Scouarnec(2015)}]{10.14778/2856318.2856324}
Andr\'{e}, F.; Kermarrec, A.-M.; and Le~Scouarnec, N. 2015.
\newblock Cache locality is not enough: high-performance nearest neighbor search with product quantization fast scan.
\newblock \emph{Proc. VLDB Endow.}, 9(4): 288–299.

\bibitem[{Anthropic(2024)}]{anthropic2024claude}
Anthropic, A. 2024.
\newblock The claude 3 model family: Opus, sonnet, haiku.
\newblock \emph{Claude-3 Model Card}.

\bibitem[{Bajaj et~al.(2016)Bajaj, Campos, Craswell, Deng, Gao, Liu, Majumder, McNamara, Mitra, Nguyen et~al.}]{bajaj2016ms}
Bajaj, P.; Campos, D.; Craswell, N.; Deng, L.; Gao, J.; Liu, X.; Majumder, R.; McNamara, A.; Mitra, B.; Nguyen, T.; et~al. 2016.
\newblock Ms marco: A human generated machine reading comprehension dataset.
\newblock \emph{arXiv preprint arXiv:1611.09268}.

\bibitem[{Borgeaud et~al.(2022)Borgeaud, Mensch, Hoffmann, Cai, Rutherford, Millican, Van Den~Driessche, Lespiau, Damoc, Clark et~al.}]{borgeaud2022improving}
Borgeaud, S.; Mensch, A.; Hoffmann, J.; Cai, T.; Rutherford, E.; Millican, K.; Van Den~Driessche, G.~B.; Lespiau, J.-B.; Damoc, B.; Clark, A.; et~al. 2022.
\newblock Improving language models by retrieving from trillions of tokens.
\newblock In \emph{International conference on machine learning}, 2206--2240. PMLR.

\bibitem[{Brown et~al.(2020)Brown, Mann, Ryder, Subbiah, Kaplan, Dhariwal, Neelakantan, Shyam, Sastry, Askell et~al.}]{brown2020language}
Brown, T.; Mann, B.; Ryder, N.; Subbiah, M.; Kaplan, J.~D.; Dhariwal, P.; Neelakantan, A.; Shyam, P.; Sastry, G.; Askell, A.; et~al. 2020.
\newblock Language models are few-shot learners.
\newblock \emph{Advances in neural information processing systems}, 33: 1877--1901.

\bibitem[{Chowdhery et~al.(2022)Chowdhery, Narang, Devlin, Bosma, Mishra, Roberts, Barham, Chung, Sutton, Gehrmann et~al.}]{chowdhery2022palm}
Chowdhery, A.; Narang, S.; Devlin, J.; Bosma, M.; Mishra, G.; Roberts, A.; Barham, P.; Chung, H.~W.; Sutton, C.; Gehrmann, S.; et~al. 2022.
\newblock PaLM: Scaling Language Modeling with Pathways.
\newblock \emph{arXiv preprint arXiv:2204.02311}.

\bibitem[{Craig(1956)}]{craig1956directed}
Craig, R.~C. 1956.
\newblock Directed versus independent discovery of established relations.
\newblock \emph{Journal of Educational Psychology}, 47(4): 223.

\bibitem[{Guu et~al.(2020)Guu, Lee, Tung, Pasupat, and Chang}]{pmlr-v119-guu20a}
Guu, K.; Lee, K.; Tung, Z.; Pasupat, P.; and Chang, M. 2020.
\newblock Retrieval Augmented Language Model Pre-Training.
\newblock In III, H.~D.; and Singh, A., eds., \emph{Proceedings of the 37th International Conference on Machine Learning}, volume 119 of \emph{Proceedings of Machine Learning Research}, 3929--3938. PMLR.

\bibitem[{Hendrycks et~al.(2020)Hendrycks, Burns, Basart, Zou, Mazeika, Song, and Steinhardt}]{hendrycks2020measuring}
Hendrycks, D.; Burns, C.; Basart, S.; Zou, A.; Mazeika, M.; Song, D.; and Steinhardt, J. 2020.
\newblock Measuring massive multitask language understanding.
\newblock \emph{arXiv preprint arXiv:2009.03300}.

\bibitem[{Izacard et~al.(2021)Izacard, Caron, Hosseini, Riedel, Bojanowski, Joulin, and Grave}]{izacard2021unsupervised}
Izacard, G.; Caron, M.; Hosseini, L.; Riedel, S.; Bojanowski, P.; Joulin, A.; and Grave, E. 2021.
\newblock Unsupervised dense information retrieval with contrastive learning.
\newblock \emph{arXiv preprint arXiv:2112.09118}.

\bibitem[{Izacard and Grave(2021)}]{izacard2021leveraging}
Izacard, G.; and Grave, E. 2021.
\newblock Leveraging Passage Retrieval with Generative Models for Open Domain Question Answering.
\newblock In \emph{EACL 2021-16th Conference of the European Chapter of the Association for Computational Linguistics}, 874--880. Association for Computational Linguistics.

\bibitem[{Izacard et~al.(2023)Izacard, Lewis, Lomeli, Hosseini, Petroni, Schick, Dwivedi-Yu, Joulin, Riedel, and Grave}]{izacard2023atlas}
Izacard, G.; Lewis, P.; Lomeli, M.; Hosseini, L.; Petroni, F.; Schick, T.; Dwivedi-Yu, J.; Joulin, A.; Riedel, S.; and Grave, E. 2023.
\newblock Atlas: Few-shot learning with retrieval augmented language models.
\newblock \emph{Journal of Machine Learning Research}, 24(251): 1--43.

\bibitem[{J{\"a}rvelin and Kek{\"a}l{\"a}inen(2002)}]{jarvelin2002cumulated}
J{\"a}rvelin, K.; and Kek{\"a}l{\"a}inen, J. 2002.
\newblock Cumulated gain-based evaluation of IR techniques.
\newblock \emph{ACM Transactions on Information Systems (TOIS)}, 20(4): 422--446.

\bibitem[{Ji et~al.(2023)Ji, Lee, Frieske, Yu, Su, Xu, Ishii, Bang, Madotto, and Fung}]{ji2023survey}
Ji, Z.; Lee, N.; Frieske, R.; Yu, T.; Su, D.; Xu, Y.; Ishii, E.; Bang, Y.~J.; Madotto, A.; and Fung, P. 2023.
\newblock Survey of hallucination in natural language generation.
\newblock \emph{ACM Computing Surveys}, 55(12): 1--38.

\bibitem[{Johnson, Douze, and J{\'e}gou(2019)}]{johnson2019billion}
Johnson, J.; Douze, M.; and J{\'e}gou, H. 2019.
\newblock Billion-scale similarity search with GPUs.
\newblock \emph{IEEE Transactions on Big Data}, 7(3): 535--547.

\bibitem[{Kandpal et~al.(2023)Kandpal, Deng, Roberts, Wallace, and Raffel}]{kandpal2023large}
Kandpal, N.; Deng, H.; Roberts, A.; Wallace, E.; and Raffel, C. 2023.
\newblock Large language models struggle to learn long-tail knowledge.
\newblock In \emph{International Conference on Machine Learning}, 15696--15707. PMLR.

\bibitem[{Karpukhin et~al.(2020{\natexlab{a}})Karpukhin, O{\u{g}}uz, Min, Lewis, Wu, Edunov, Chen, and Yih}]{karpukhin2020dense}
Karpukhin, V.; O{\u{g}}uz, B.; Min, S.; Lewis, P.; Wu, L.; Edunov, S.; Chen, D.; and Yih, W.-t. 2020{\natexlab{a}}.
\newblock Dense passage retrieval for open-domain question answering.
\newblock \emph{arXiv preprint arXiv:2004.04906}.

\bibitem[{Karpukhin et~al.(2020{\natexlab{b}})Karpukhin, Oguz, Min, Lewis, Wu, Edunov, Chen, and Yih}]{karpukhin-etal-2020-dense}
Karpukhin, V.; Oguz, B.; Min, S.; Lewis, P.; Wu, L.; Edunov, S.; Chen, D.; and Yih, W.-t. 2020{\natexlab{b}}.
\newblock Dense Passage Retrieval for Open-Domain Question Answering.
\newblock In \emph{Proceedings of the 2020 Conference on Empirical Methods in Natural Language Processing (EMNLP)}, 6769--6781. Online: Association for Computational Linguistics.

\bibitem[{Kittell(1957)}]{kittell1957experimental}
Kittell, J.~E. 1957.
\newblock An experimental study of the effect of external direction during learning on transfer and retention of principles.
\newblock \emph{Journal of Educational Psychology}, 48(7): 391.

\bibitem[{Kojima et~al.(2022)Kojima, Gu, Reid, Matsuo, and Iwasawa}]{kojima2022large}
Kojima, T.; Gu, S.~S.; Reid, M.; Matsuo, Y.; and Iwasawa, Y. 2022.
\newblock Large language models are zero-shot reasoners.
\newblock \emph{Advances in neural information processing systems}, 35: 22199--22213.

\bibitem[{Kwiatkowski et~al.(2019)Kwiatkowski, Palomaki, Redfield, Collins, Parikh, Alberti, Epstein, Polosukhin, Devlin, Lee, Toutanova, Jones, Kelcey, Chang, Dai, Uszkoreit, Le, and Petrov}]{10.1162/tacl_a_00276}
Kwiatkowski, T.; Palomaki, J.; Redfield, O.; Collins, M.; Parikh, A.; Alberti, C.; Epstein, D.; Polosukhin, I.; Devlin, J.; Lee, K.; Toutanova, K.; Jones, L.; Kelcey, M.; Chang, M.-W.; Dai, A.~M.; Uszkoreit, J.; Le, Q.; and Petrov, S. 2019.
\newblock {Natural Questions: A Benchmark for Question Answering Research}.
\newblock \emph{Transactions of the Association for Computational Linguistics}, 7: 453--466.

\bibitem[{Lee et~al.(2023)Lee, Phatale, Mansoor, Lu, Mesnard, Bishop, Carbune, and Rastogi}]{lee2023rlaif}
Lee, H.; Phatale, S.; Mansoor, H.; Lu, K.; Mesnard, T.; Bishop, C.; Carbune, V.; and Rastogi, A. 2023.
\newblock Rlaif: Scaling reinforcement learning from human feedback with ai feedback.
\newblock \emph{arXiv preprint arXiv:2309.00267}.

\bibitem[{Lee, Chang, and Toutanova(2019)}]{lee2019latent}
Lee, K.; Chang, M.-W.; and Toutanova, K. 2019.
\newblock Latent Retrieval for Weakly Supervised Open Domain Question Answering.
\newblock In \emph{Proceedings of the 57th Annual Meeting of the Association for Computational Linguistics}, 6086--6096.

\bibitem[{Lewis et~al.(2020)Lewis, Perez, Piktus, Petroni, Karpukhin, Goyal, K{\"u}ttler, Lewis, Yih, Rockt{\"a}schel et~al.}]{lewis2020retrieval}
Lewis, P.; Perez, E.; Piktus, A.; Petroni, F.; Karpukhin, V.; Goyal, N.; K{\"u}ttler, H.; Lewis, M.; Yih, W.-t.; Rockt{\"a}schel, T.; et~al. 2020.
\newblock Retrieval-augmented generation for knowledge-intensive nlp tasks.
\newblock \emph{Advances in Neural Information Processing Systems}, 33: 9459--9474.

\bibitem[{Lin et~al.(2023)Lin, Chen, Chen, Shi, Lomeli, James, Rodriguez, Kahn, Szilvasy, Lewis et~al.}]{lin2023ra}
Lin, X.~V.; Chen, X.; Chen, M.; Shi, W.; Lomeli, M.; James, R.; Rodriguez, P.; Kahn, J.; Szilvasy, G.; Lewis, M.; et~al. 2023.
\newblock Ra-dit: Retrieval-augmented dual instruction tuning.
\newblock \emph{arXiv preprint arXiv:2310.01352}.

\bibitem[{Mallen et~al.(2022)Mallen, Asai, Zhong, Das, Khashabi, and Hajishirzi}]{mallen2022not}
Mallen, A.; Asai, A.; Zhong, V.; Das, R.; Khashabi, D.; and Hajishirzi, H. 2022.
\newblock When not to trust language models: Investigating effectiveness of parametric and non-parametric memories.
\newblock \emph{arXiv preprint arXiv:2212.10511}.

\bibitem[{Mayer(2004)}]{mayer2004should}
Mayer, R.~E. 2004.
\newblock Should there be a three-strikes rule against pure discovery learning?
\newblock \emph{American psychologist}, 59(1): 14.

\bibitem[{Ouyang et~al.(2022)Ouyang, Wu, Jiang, Almeida, Wainwright, Mishkin, Zhang, Agarwal, Slama, Ray et~al.}]{ouyang2022training}
Ouyang, L.; Wu, J.; Jiang, X.; Almeida, D.; Wainwright, C.; Mishkin, P.; Zhang, C.; Agarwal, S.; Slama, K.; Ray, A.; et~al. 2022.
\newblock Training language models to follow instructions with human feedback.
\newblock \emph{Advances in neural information processing systems}, 35: 27730--27744.

\bibitem[{Petroni et~al.(2021)Petroni, Piktus, Fan, Lewis, Yazdani, De~Cao, Thorne, Jernite, Karpukhin, Maillard, Plachouras, Rockt{\"a}schel, and Riedel}]{petroni-etal-2021-kilt}
Petroni, F.; Piktus, A.; Fan, A.; Lewis, P.; Yazdani, M.; De~Cao, N.; Thorne, J.; Jernite, Y.; Karpukhin, V.; Maillard, J.; Plachouras, V.; Rockt{\"a}schel, T.; and Riedel, S. 2021.
\newblock {KILT}: a Benchmark for Knowledge Intensive Language Tasks.
\newblock In \emph{Proceedings of the 2021 Conference of the North American Chapter of the Association for Computational Linguistics: Human Language Technologies}, 2523--2544. Online: Association for Computational Linguistics.

\bibitem[{Piaget(1970)}]{piaget1970science}
Piaget, J. 1970.
\newblock \emph{Science of Education and the Psychology of the Child}.
\newblock Orion Press.
\newblock ISBN 9780670621729.

\bibitem[{Ram et~al.(2023)Ram, Levine, Dalmedigos, Muhlgay, Shashua, Leyton-Brown, and Shoham}]{10.1162/tacl_a_00605}
Ram, O.; Levine, Y.; Dalmedigos, I.; Muhlgay, D.; Shashua, A.; Leyton-Brown, K.; and Shoham, Y. 2023.
\newblock {In-Context Retrieval-Augmented Language Models}.
\newblock \emph{Transactions of the Association for Computational Linguistics}, 11: 1316--1331.

\bibitem[{Reimers and Gurevych(2019{\natexlab{a}})}]{reimers2019sentence}
Reimers, N.; and Gurevych, I. 2019{\natexlab{a}}.
\newblock Sentence-BERT: Sentence Embeddings using Siamese BERT-Networks.
\newblock In \emph{Proceedings of the 2019 Conference on Empirical Methods in Natural Language Processing and the 9th International Joint Conference on Natural Language Processing (EMNLP-IJCNLP)}, 3982--3992.

\bibitem[{Reimers and Gurevych(2019{\natexlab{b}})}]{reimers-2019-sentence-bert}
Reimers, N.; and Gurevych, I. 2019{\natexlab{b}}.
\newblock Sentence-BERT: Sentence Embeddings using Siamese BERT-Networks.
\newblock In \emph{Proceedings of the 2019 Conference on Empirical Methods in Natural Language Processing}. Association for Computational Linguistics.

\bibitem[{Robertson, Zaragoza et~al.(2009)}]{robertson2009probabilistic}
Robertson, S.; Zaragoza, H.; et~al. 2009.
\newblock The probabilistic relevance framework: BM25 and beyond.
\newblock \emph{Foundations and Trends{\textregistered} in Information Retrieval}, 3(4): 333--389.

\bibitem[{Salemi et~al.(2023)Salemi, Mysore, Bendersky, and Zamani}]{salemi2023lamp}
Salemi, A.; Mysore, S.; Bendersky, M.; and Zamani, H. 2023.
\newblock La{MP}: When Large Language Models Meet Personalization.
\newblock arXiv:2304.11406.

\bibitem[{Shi et~al.(2023)Shi, Min, Yasunaga, Seo, James, Lewis, Zettlemoyer, and Yih}]{shi2023replug}
Shi, W.; Min, S.; Yasunaga, M.; Seo, M.; James, R.; Lewis, M.; Zettlemoyer, L.; and Yih, W.-t. 2023.
\newblock Replug: Retrieval-augmented black-box language models.
\newblock \emph{arXiv preprint arXiv:2301.12652}.

\bibitem[{Shulman et~al.(1966)Shulman, Keislar, University, on~Learning, and the Educational~Process}]{shulman1966learning}
Shulman, L.; Keislar, E.; University, S.; on~Learning, S. S. R. C. U.~C.; and the Educational~Process. 1966.
\newblock \emph{Learning by Discovery: A Critical Appraisal}.
\newblock Rand McNally education series. Rand McNally.

\bibitem[{Thorne et~al.(2018)Thorne, Vlachos, Christodoulopoulos, and Mittal}]{Thorne18Fever}
Thorne, J.; Vlachos, A.; Christodoulopoulos, C.; and Mittal, A. 2018.
\newblock {FEVER}: a Large-scale Dataset for Fact Extraction and {VERification}.
\newblock In \emph{NAACL-HLT}.

\bibitem[{Wang, Yang, and Wei(2023)}]{wang2023learning}
Wang, L.; Yang, N.; and Wei, F. 2023.
\newblock Learning to retrieve in-context examples for large language models.
\newblock \emph{arXiv preprint arXiv:2307.07164}.

\bibitem[{Wang et~al.(2022)Wang, Kordi, Mishra, Liu, Smith, Khashabi, and Hajishirzi}]{wang2022self}
Wang, Y.; Kordi, Y.; Mishra, S.; Liu, A.; Smith, N.~A.; Khashabi, D.; and Hajishirzi, H. 2022.
\newblock Self-instruct: Aligning language models with self-generated instructions.
\newblock \emph{arXiv preprint arXiv:2212.10560}.

\bibitem[{Wei et~al.(2022)Wei, Wang, Schuurmans, Bosma, Xia, Chi, Le, Zhou et~al.}]{wei2022chain}
Wei, J.; Wang, X.; Schuurmans, D.; Bosma, M.; Xia, F.; Chi, E.; Le, Q.~V.; Zhou, D.; et~al. 2022.
\newblock Chain-of-thought prompting elicits reasoning in large language models.
\newblock \emph{Advances in neural information processing systems}, 35: 24824--24837.

\bibitem[{Xiao and Liu(2023)}]{xiao2023bge}
Xiao, S.; and Liu, Z. 2023.
\newblock BAAI General Embedding.

\bibitem[{Xiong et~al.(2020)Xiong, Xiong, Li, Tang, Liu, Bennett, Ahmed, and Overwijk}]{xiong2020approximate}
Xiong, L.; Xiong, C.; Li, Y.; Tang, K.-F.; Liu, J.; Bennett, P.; Ahmed, J.; and Overwijk, A. 2020.
\newblock Approximate nearest neighbor negative contrastive learning for dense text retrieval.
\newblock \emph{arXiv preprint arXiv:2007.00808}.

\bibitem[{Yang et~al.(2018)Yang, Qi, Zhang, Bengio, Cohen, Salakhutdinov, and Manning}]{yang2018hotpotqa}
Yang, Z.; Qi, P.; Zhang, S.; Bengio, Y.; Cohen, W.; Salakhutdinov, R.; and Manning, C.~D. 2018.
\newblock HotpotQA: A Dataset for Diverse, Explainable Multi-hop Question Answering.
\newblock In \emph{Proceedings of the 2018 Conference on Empirical Methods in Natural Language Processing}. Association for Computational Linguistics.

\bibitem[{Yu et~al.(2023)Yu, Xiong, Yu, and Liu}]{yu2023augmentation}
Yu, Z.; Xiong, C.; Yu, S.; and Liu, Z. 2023.
\newblock Augmentation-adapted retriever improves generalization of language models as generic plug-in.
\newblock \emph{arXiv preprint arXiv:2305.17331}.

\bibitem[{Zhang et~al.(2023{\natexlab{a}})Zhang, Xiao, Liu, Dou, and Nie}]{zhang2023retrieve}
Zhang, P.; Xiao, S.; Liu, Z.; Dou, Z.; and Nie, J.-Y. 2023{\natexlab{a}}.
\newblock Retrieve anything to augment large language models.
\newblock \emph{arXiv preprint arXiv:2310.07554}.

\bibitem[{Zhang et~al.(2023{\natexlab{b}})Zhang, Li, Cui, Cai, Liu, Fu, Huang, Zhao, Zhang, Chen et~al.}]{zhang2023siren}
Zhang, Y.; Li, Y.; Cui, L.; Cai, D.; Liu, L.; Fu, T.; Huang, X.; Zhao, E.; Zhang, Y.; Chen, Y.; et~al. 2023{\natexlab{b}}.
\newblock Siren's song in the AI ocean: a survey on hallucination in large language models.
\newblock \emph{arXiv preprint arXiv:2309.01219}.

\end{thebibliography}

\section{Technical Appendix}

\subsection{Out-of-Domain Task Validation}

\begin{table*}[ht]
\centering
\small
\begin{tabular}{lcccccccccccc}
\toprule
\multirow{2}{*}{\textbf{Method}} & \textbf{LaMP 1U} & \multicolumn{2}{c}{\textbf{LaMP 2U}} & \multicolumn{2}{c}{\textbf{LaMP 3U}} & \multicolumn{2}{c}{\textbf{LaMP 4U}} & \multicolumn{2}{c}{\textbf{LaMP 5U}} & \multicolumn{2}{c}{\textbf{LaMP 7U}} \\
\cmidrule(lr){2-2} \cmidrule(lr){3-4} \cmidrule(lr){5-6} \cmidrule(lr){7-8} \cmidrule(lr){9-10} \cmidrule(lr){11-12}
 & Acc & Acc & F1 & MAE & MSE & Rouge-1 & Rouge-L & Rouge-1 & Rouge-L & Rouge-1 & Rouge-L \\
\midrule
No retrieval & 52.9 & 43.4 & 0.367 & 0.369 & 0.660 & 0.136 & 0.118 & 0.390 & 0.316 & 0.440 & \textbf{0.394} \\
Contriever & 74.5 & 50.0 & 0.450 & 0.307 & 0.599 & 0.163 & 0.147 & 0.396 & 0.343 & \textbf{0.442} & 0.390 \\
FiGRet-Contriever & \textbf{75.3} & \textbf{50.9} & \textbf{0.458} & \textbf{0.299} & \textbf{0.579} & \textbf{0.167} & \textbf{0.150} & \textbf{0.423} & \textbf{0.362} & 0.441 & 0.391 \\
\bottomrule
\end{tabular}
\caption{Performance comparison on LaMP benchmark tasks}
\label{tab:lamp_results}
\end{table*}

To validate the generalizability of our proposed framework on tasks unrelated to the upstream training objective, we evaluated its performance on the LaMP (Language Models Personalization, \citet{salemi2023lamp}) benchmark. LaMP is a publicly available evaluation dataset for personalized language modeling, designed to assess various aspects of personalization.

Following the original experimental setup of the LaMP benchmark, we employed GPT-3.5-Turbo as the reasoning LLM and Contriever as the retrieval model. We evaluated the performance on various subtasks both before and after applying our framework (excluding task 6, "Email Subject Generation", due to data accessibility issues).

Table~\ref{tab:lamp_results} presents the results. Our framework demonstrates consistent performance improvements across six diverse tasks: Citation Identification, Movie Tagging, Product Rating, News Headline Generation, Scholarly Title Generation, and Tweet Paraphrasing. The only exception is LaMP 7, where both our framework and the baseline Contriever model led to performance degradation. This suggests that the task might benefit minimally from retrieved documents.

These results indicate that our proposed framework can generalize to enhance performance on out-of-domain tasks. We hypothesize that this stems from the pedagogical approach of our framework, which shifts the retriever's focus from "relevance-based retrieval" to "LLM generation preference-aware retrieval." This allows the retriever to provide knowledge to the LLM in a more fundamentally aligned manner. 

\subsection{Performance Across Different Numbers of Documents}

\begin{figure}[b]
\centering
\includegraphics[width=0.35\textwidth]{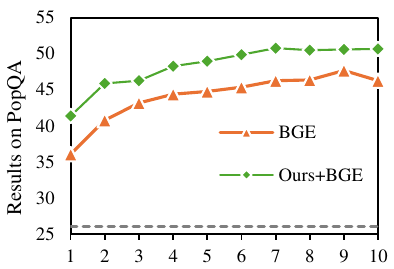}
\caption{
Aspect-wise Retrieval Improvements After Training.
}
\label{fig: doc_num}
\vspace{-10pt}
\end{figure}

To observe the performance variations of our method across different numbers of documents, we conducted experiments on the PopQA, comparing the performance of the BGE model before and after applying our framework, using Llama3-8B-Instruct as the inference LLM. The results, as shown in Figure~\ref{fig: doc_num}, indicate that our framework consistently achieves stable performance improvements across various document counts.

\subsection{Showcase of Constructing Guidance}

\begin{figure*}[ht]
\centering
\includegraphics[width=\textwidth]{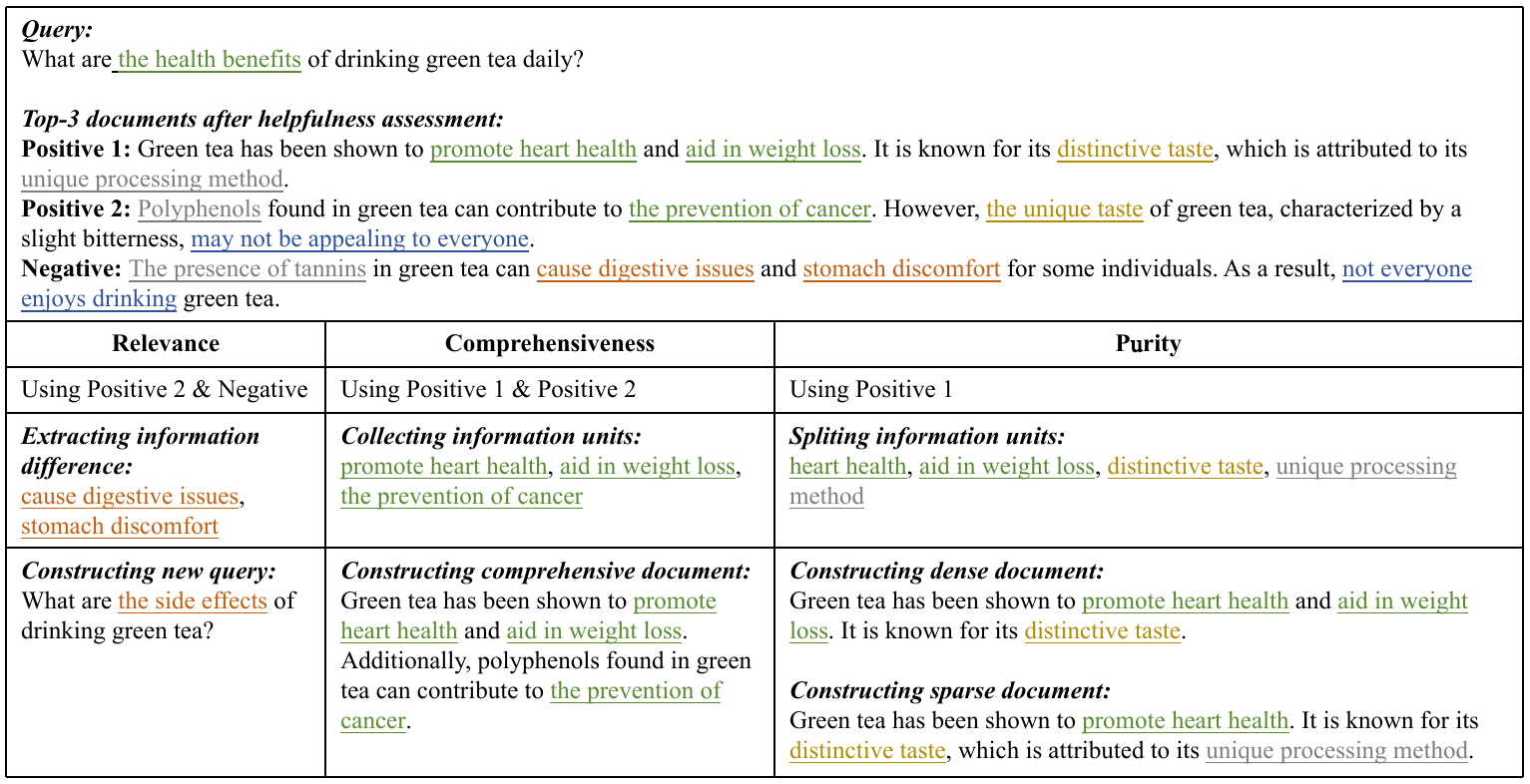}
\vspace{-10pt}
\caption{
Showcase of Guidance Example Construction.
}
\label{fig:showcase}
\vspace{-10pt}
\end{figure*}

We present a showcase of constructing guidance examples as shown in Figure~\ref{fig:showcase}, corresponding to the content in Figure 1 of the main paper.

\subsection{Prompt Used in Constructing Guidance}
\begin{table*}[ht]
\small
\begin{tabularx}{\textwidth}{m{15cm}}
\toprule
\textsc{Teacher Model Scoring} \\
\midrule
\textbf{Instruction:} \\
- Compare the provided passages in terms of their helpfulness in answering the provided query.\\
- Rate each passage's helpfulness based on relevance, comprehensiveness, and proportion of useful information.\\
- Relevance is the primary factor determining the score range:\\
\quad - 8 to 10: Directly relevant, containing information that can directly answer the query.\\
\quad - 4 to 7: Possibly or indirectly relevant, providing some useful information but not directly answering the query.\\
\quad - 0 to 3: Not containing potentially relevant information.\\
- Within each range, determine the specific score based on comprehensiveness and proportion of useful information:\\
\quad - High comprehensiveness and high proportion of useful information: Highest score in the range.\\
\quad - Low comprehensiveness and low proportion of useful information: Lowest score in the range.\\
\quad - Other cases: Consider both factors to assign a middle score.\\
\\
\textbf{Query:} \\
\{query\} \\
\\
\textbf{Documents:} \\
\{documents with IDs\} \\
\bottomrule
\end{tabularx}
\caption{Prompt for Teacher Model Scoring.}
\label{tab:prompt_helpfulness}
\end{table*}

\begin{table*}[ht]
\small
\begin{tabularx}{\textwidth}{m{15cm}}
\toprule
\textsc{Relevance (Part I)} \\
\midrule
(Continuing the conversation from Teacher Model Scoring)\\
\\
\textbf{Instruction:} \\
- Identify and summarize the content differences between documents that are less helpful and more helpful in answering the query. \\
\quad - Highlight the unique content elements present in less helpful documents and absent in more helpful documents. \\
- List the IDs of the documents deemed less helpful based on these differences. \\
- Respond with these differences and their corresponding document IDs in JSON format. \\
\midrule
\textsc{Relevance (Part II)} \\
\midrule
(Continuing the conversation from Part I)\\
\\
\textbf{Instruction:} \\
- For each identified content difference, slightly modify the original query to create a different, new, and concise query that preserves the original subject. \\
\quad - The modification should make the corresponding documents highly relevant in answering this new query, while the documents previously helpful for the original query should become less relevant. \\
- List the documents that can now effectively answer the new query (positive documents) and those that cannot (negative documents), considering all the documents. \\
- Respond with the new queries and the corresponding positive document IDs and negative document IDs in JSON format. \\
\bottomrule
\end{tabularx}
\caption{Prompts for Constructing Examples in the Relevance Objective.}
\label{tab:prompt_relevance}
\end{table*}

\begin{table*}[ht]
\small
\begin{tabularx}{\textwidth}{m{15cm}}
\toprule
\textsc{Comprehensive} \\
\midrule
\textbf{Instruction:} \\
- Given a query and several documents, please extract all the information units highly relevant to the query and compose a new, complete, and concise document. \\
- Avoid including irrelevant details to keep the document short and concise. \\
- Respond with the new document in JSON format. \\
\\
\textbf{Query:} \\
\{query\} \\
\\
\textbf{Documents:} \\
\{helpful documents\} \\
\bottomrule
\end{tabularx}
\caption{Prompt for Constructing Examples in the Comprehensiveness Objective.}
\label{tab:prompt_comprehensiveness}
\end{table*}

\begin{table*}[ht]
\small
\begin{tabularx}{\textwidth}{m{15cm}}
\toprule
\textsc{Purity} \\
\midrule
\textbf{Instruction:} \\
- Given a query and a document, please selectively remove some information from the document that is related and unrelated to the query to form two new documents. One should contain more (dense document) and the other less (sparse document) relevant information. \\
- Each new document should be as complete as possible, forming independent and coherent documents. \\
- Respond sequentially with the dense document and the sparse document in JSON format. \\
\\
\textbf{Query:} \\
\{query\} \\
\\
\textbf{Document:} \\
\{document\} \\
\bottomrule
\end{tabularx}
\caption{Prompt for Constructing Examples in the Purity Objective.}
\label{tab:prompt_information_density}
\end{table*}

The detailed contents of the prompts used in constructing guidance across different objectives of document analysis are provided in the following tables. 
Table~\ref{tab:prompt_helpfulness} provides prompt for teacher model scoring. 
Table~\ref{tab:prompt_relevance} provides prompts for constructing examples in the relevance objective. 
Table~\ref{tab:prompt_comprehensiveness} provides prompt for constructing examples in the comprehensiveness objective.
Table~\ref{tab:prompt_information_density} provides prompt for constructing examples in the purity objective.

\subsection{Prompt Used in Inference}
\begin{table*}[ht]
\small
\begin{tabularx}{\textwidth}{m{15cm}}
\toprule
\textsc{Prompt for MMLU} \\
\midrule
\textbf{Knowledge:} \\
\{documents\} \\
\\
\{Few-shot Examples\} \\
\\
\textbf{Question:} \{question\} \\
A: \{choice A\} \\
B: \{choice B\} \\
C: \{choice C\} \\
D: \{choice D\} \\
\textbf{Answer:} \\
\bottomrule
\end{tabularx}
\caption{Prompt for MMLU.}
\label{tab:prompt_MMLU}
\end{table*}

\begin{table*}[ht]
\small
\begin{tabularx}{\textwidth}{m{15cm}}
\toprule
\textsc{Prompt for Open-domain QA} \\
\midrule
\textbf{Knowledge:} \\
\{documents\} \\
\\
Provide direct answers to the following questions: \\
\\
\{Few-shot Examples\} \\
\\
\textbf{Question:} \{question\} \\
\bottomrule
\end{tabularx}
\caption{Prompt for Open-domain QA.}
\label{tab:prompt_QA}
\end{table*}

\begin{table*}[ht]
\small
\begin{tabularx}{\textwidth}{m{15cm}}
\toprule
\textsc{Prompt for Fact Checking} \\
\midrule
\textbf{Knowledge:} \\
\{documents\} \\
\\
Evaluate the following statements and respond with "SUPPORTS" if you agree or "REFUTES" if you disagree: \\
\\
\{Few-shot Examples\} \\
\\
\textbf{Question:} \{question\} \\
\bottomrule
\end{tabularx}
\caption{Prompt for Fact Checking.}
\label{tab:prompt_FC}
\end{table*}

The detailed contents of the prompts used for LLMs during inference in our experiments are provided in the following tables.
Table~\ref{tab:prompt_MMLU} provides the prompt utilized for the MMLU dataset.
Table~\ref{tab:prompt_QA} provides the prompt utilized for the open-domain question answering task.
Table~\ref{tab:prompt_FC} provides the prompt utilized for the fact checking task.

\subsection{Hyperparameter Settings}

\paragraph{Curriculum Learning}
The temperatures $T_1$ and $T_2$ were set to 2 and 0.2, respectively.

\paragraph{Training Procedure}
We trained our model using the following hyperparameters:

\begin{itemize}
    \item \textbf{Hardware:} 4 NVIDIA A100 GPUs
    \item \textbf{Batch size:} 32 per GPU
    \item \textbf{Training samples:} Each training sample consisted of a query, a positive document, and a negative document, randomly sampled from the set of guidance examples.
    \item \textbf{Epochs:} 1
    \item \textbf{Learning rate scheduler:} Cosine
    \item \textbf{Weight decay:} 0.1
    \item \textbf{Pooling:} Following the original models' training procedures, we used CLS pooling for the BGE model and average pooling for both the Contriever and SBERT models.
    \item \textbf{Similarity Calculation:}  We adopted cosine similarity for the Contriever model and dot product similarity for both the BGE and SBERT models, consistent with their original implementations. 
\end{itemize} 

\end{document}